\documentclass[aps, amsmath,amssymb, preprint,superscriptaddress]{revtex4-1}%
\usepackage{graphicx}

\usepackage{amsmath}
\usepackage{amsfonts}
\usepackage{amssymb}

\usepackage{bm}
\usepackage{sidecap}

\usepackage{epstopdf}
\usepackage{hyperref}
\usepackage{color}

\providecommand{\U}[1]{\protect\rule{.1in}{.1in}}
\begin{document}

\title{Tunable mid-infrared coherent perfect absorption\\ in a graphene meta-surface}

\author{Yuancheng Fan}
\email{phyfan@nwpu.edu.cn}
\affiliation{Key Laboratory of Space Applied Physics and Chemistry, Ministry of Education and Department of Applied Physics, School of Science, Northwestern Polytechnical University, Xi'an 710129, China}
\author{Zhe Liu}
\affiliation{Beijing National Laboratory for Condensed Matter Physics, Institute of Physics, Chinese Academy of Sciences, Beijing 100190, China}
\author{Fuli Zhang}
\affiliation{Key Laboratory of Space Applied Physics and Chemistry, Ministry of Education and Department of Applied Physics, School of Science, Northwestern Polytechnical University, Xi'an 710129, China}
\author{Qian Zhao}
\affiliation{State Key Laboratory of Tribology, Department of Mechanical Engineering, Tsinghua University, Beijing 100084, China}
\author{Zeyong Wei}
\affiliation{Key Laboratory of Advanced Micro-structure Materials (MOE) and School of Physics Science and Engineering, Tongji University, Shanghai 200092, China}
\author{Quanhong Fu}
\affiliation{Key Laboratory of Space Applied Physics and Chemistry, Ministry of Education and Department of Applied Physics, School of Science, Northwestern Polytechnical University, Xi'an 710129, China}
\author{Junjie Li}
\affiliation{Beijing National Laboratory for Condensed Matter Physics, Institute of Physics, Chinese Academy of Sciences, Beijing 100190, China}
\author{Changzhi Gu}
\affiliation{Beijing National Laboratory for Condensed Matter Physics, Institute of Physics, Chinese Academy of Sciences, Beijing 100190, China}
\author{Hongqiang Li}
\affiliation{Key Laboratory of Advanced Micro-structure Materials (MOE) and School of Physics Science and Engineering, Tongji University, Shanghai 200092, China}

\begin{abstract}
We exploited graphene nanoribbons based meta-surface to realize coherent perfect absorption (CPA) in the mid-infrared regime. It was shown that quasi-CPA frequencies, at which CPA can be demonstrated with proper phase modulations, exist for the graphene meta-surface with strong resonant behaviors. The CPA can be tuned substantially by merging the geometric design of the meta-surface and the electrical tunability of graphene. Furthermore, we found that the graphene nanoribbon meta-surface based CPA is realizable with experimental graphene data. The findings of CPA with graphene meta-surface can be generalized for potential applications in optical detections and signal processing with two-dimensional optoelectronic materials.
\end{abstract}

\maketitle

Graphene has recently attracted considerable attention for both its interesting physics and potential applications. The realistic two-dimensional (2D) material is merging in photonics and optoelectronics.\cite{Bonaccorso2010,Yan2012,Abajo2014,Low2014,Tassin2013} It exhibits much stronger binding of surface plasmon polaritons and supports its relatively longer propagation.\cite{Jablan2009} Linear dispersion of the 2D Dirac fermions provides ultrawideband tunability through electrostatic field, magnetic field or chemical doping.\cite{Koppens2011,Engheta2011} Graphene is almost transparent to optical waves,\cite{Nair2008} which is one remarkable feature of the two-dimensional material, and it is due to the relatively low carrier concentrations in a monolayer atomic sheet. Boosting the light-matter interaction in graphene is one important issue to address to take further advantage of graphene in optoelectronic devices. For example, to realize functionality like optical insulator \cite{Yablonovitch2001} similar to gapped graphene \cite{Yao2007,Fan2012} for nanoelectronics, which is essential for myriad applications in all-optical systems and components of much miniaturized optical circuits.\cite{Engheta2007,Fan2011}

A kind of artificial composite consisting of subwavelength-sized resonant building blocks as their `elementary particles' (meta-atoms)---metamaterial,\cite{Smith2004, Soukoulis2011} has been employed as a platform for enhancing light-matter interactions \cite{Hess2012,Kauranen2012} in graphene.\cite{Koppens2011} Especially the metamaterials with a single planar function-layer, namely meta-surface,\cite{Kildishev2013,Yu2014} has drawn enormous attention in recent years for various possibilities to manipulate light peculiarly.\cite{Yu2011,Ni2012,Sun2012,Yin2013,Shitrit2013} The meta-surface has also found its close connection to graphene: replacing metals with graphene makes the meta-surface even miniaturized for integrated optics; and the light-graphene interactions can be significantly modified in an atomically thin graphene meta-surface in return. It was reported that optical absorption enhancement can be achieved in periodically doped graphene nanodisks, in which periodic graphene nanodisks are overlying on a substrate or on a dielectric film coating on metal, the excitation of electric mode of the nanodisks together with multi-reflection from the assistants of total internal reflection and metal reflection can result in a complete optical absorption.\cite{Thongrattanasiri2012} Graphene micro/nanoribbons, ring resonators, mantles, nano-crosses and super-lattices have also been attempted for  manipulating the terahertz/infrared waves.\cite{Ju2011,Alaee2012,He2012,Yan2013,Zhang2014,Liu2012,Wang2012,Papasimakis2013,Fan2013b,Chen2011,Cheng2013,Fan2013a,Zhu2014} A comparative study \cite{Fan2013b} found that the fundamental electric dipolar mode provides stronger light-graphene interaction at terahertz frequencies than the magnetic and higher-order electric modes, it is also found that the maximum absorption in a monolayer graphene is 50\%, which is associated with a real number effective surface conductivity $2/\eta_0$ where $\eta_0$ represents the characteristic impedance of vacuum.\cite{Fan2015} The concept of coherent perfect absorption \cite{Wan2011,Zhang2012,Lin2011,Longhi2010,Hao2011,Sun2014,Kang2014} was introduced into a suspending monolayer graphene for a 100\% absorption of terahertz wave.\cite{Fan2014} However, the proposed non-resonant CPA based on the intrinsic Drude response of graphene is realizable at only several terahertz with achievable graphene samples.

In this paper, we propose a tunable mid-infrared coherent perfect absorber with a graphene nanoribbon based meta-surface. It is found that quasi-CPA frequencies, which is the necessary formation condition of coherent absorption, does exist in the mid-infrared regime for properly designed graphene nanoribbon arrays. The scattering of coherent beams can be perfectly suppressed at the quasi-CPA frequencies with proper phase modulations on the input beams. For the case with two crosses on the transmission and reflection spectra, we can achieve coherent perfect absorption at the two quasi-CPA frequencies, simultaneously. The flexible tunabilities of the graphene meta-surface based CPA are of interests for tunable infrared detections and signal modulations.

Figure 1 shows the schematic of the proposed graphene nanoribbons based meta-surface and corresponding excitation configuration with two counter-propagating and coherently modulated optical beams ($I_+$ and $I_-$), $O_+$ and $O_-$ are the respective output magnitudes, the graphene sheet is illuminated with perpendicularly polarized light (electric vector $\textit{\textbf{E}}$ is parallel to the $x$-axis). The meta-atoms, i.e., graphene nanoribbons, are periodically arranged in $x$-$y$ plane with the lattice (along $x$ axis) constant and ribbon's width being $P$ and $w$, respectively. Both lattice constant and width of the graphene nanoribbon meta-atoms play important roles in determining the optical resonant behaviors of graphene meta-surface, the lattice constant is set to be $P = 0.7$ $\mu$m for the study of CPA in the mid-infrared regime.
\begin{figure}[t]
\includegraphics[width=7.2cm]{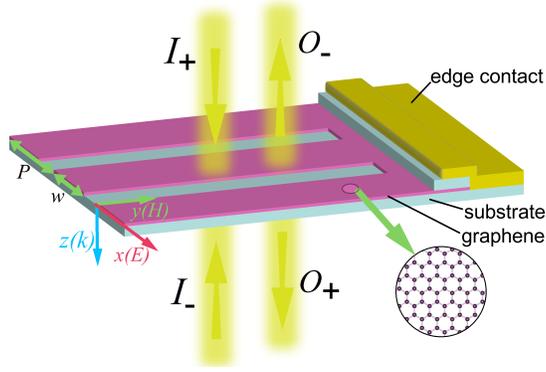}
\caption{\label{fig1}Schematic of a graphene ribbon meta-surface illustrated by two counter-propagating and coherently modulated input beams ($I_+$ and $I_-$, the electric polarization is along $x$-axis, and propagate along $\pm z$-directions), $O_+$ and $O_-$ represent the amplitudes of output beams.}
\end{figure}

In our numerical calculations, the graphene layer is considered as a sheet material modeled with complex surface conductivity ($\sigma_{g}$) since a one-atom-thick graphene sheet is sufficiently thin compared with the concerned wavelength. In the theoretical perspective based on random-phase-approximation (RPA),\cite{Wunsch2006,Hwang2007,Gusynin2007} the complex conductivity of graphene can be described by the Drude model as $\sigma_g=i e^{2}E_{F}/\pi\hbar^{2}\left(\omega+i\tau^{-1}\right)$, especially in heavily doped region and low frequencies (far below Fermi energy), where $E_F$ represents the Fermi energy, $\tau=\mu E_{F}/e\upsilon_{F}^2$ is the relaxation rate with the mobility $\mu=10^{4}$ cm$^2$V$^{-1}$s$^{-1}$ and Fermi velocity $\upsilon_{F}\approx10^{6}$m/s.
\begin{figure}[t]
\includegraphics[width=7.2cm]{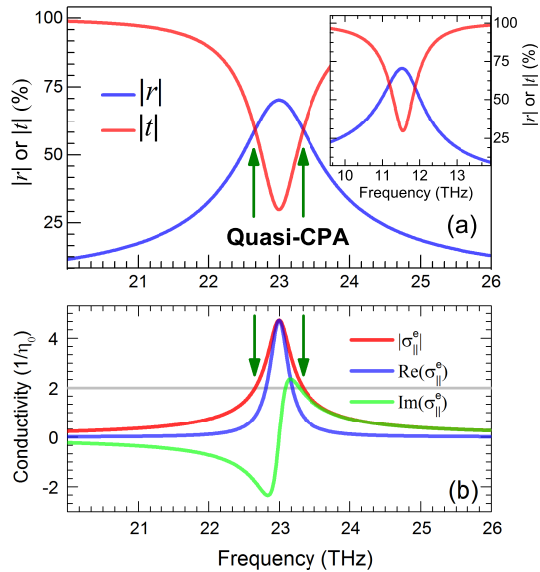}
\caption{\label{fig2}(a) Spectra of the reflection and transmission coefficients of a graphene nanoribbon meta-surface ($w =0.33$ $\mu$m), the quasi-CPA points at the crosses of the scattering spectra are indicated with arrows. (b) Absolute values, real and imaginary parts of the effective surface conductivities of the corresponding graphene nanoribbon meta-surface.}
\end{figure}

We first took $E_{F}=0.5$ eV. Figure 2(a) presents the spectra of the reflection coefficients ($r$) and transmission coefficients ($t$) for the case of graphene nanoribbon meta-surface with width of $w = 0.33$ $\mu$m. We can see an obvious resonance around $23.0$ THz. The resonance was confirmed to be electric dipolar mode (which will be verified from the effective surface conductivity below), similar to the low-frequency resonance of split-ring-resonators as that in Ref. 35. The excitation of electric dipolar mode leads to the enhancement of absorption in the graphene sheet, however, the maximum limit is 50\%.

The complex scattering coefficients ($O_\pm$) of the graphene nanoribbon meta-surface can be related to the two input beams ($I_\pm$, and in this paper the two input beams are set to be of equal amplitude $I$) through a scattering matrix, $S_g$, defined as:
\begin{equation}
\begin{pmatrix}
O_+ \\
O_-
\end{pmatrix} =S_g
\begin{pmatrix}
I_+ \\
I_-
\end{pmatrix}=
\begin{pmatrix}
t_+ & r_- \\
r_+ & t_-
\end{pmatrix}
\begin{pmatrix}
Ie^{i\phi_+} \\
Ie^{i\phi_-}
\end{pmatrix},
\end{equation}
where $(t/r)_+$ and $(t/r)_-$ are scattering elements of forward (irradiate towards $z_+$, $I_+$) and backward (irradiate towards $z_-$, $I_-$) beams, since the linear monolayer graphene under investigation is of reciprocity and spatial symmetry, the scattering matrix can be simplified with $t_\pm=t$ and $r_\pm=r$. Then the amplitude of the scattering coefficients would be
\begin{equation}
\left|O_+\right|=\left|O_-\right| =\left|tIe^{i\phi_+}+rIe^{i\phi_-}\right|.
\end{equation}
The scatterings of the input beams are required to be inhibited to demonstrate a CPA, that means $tIe^{i\phi_+}=rIe^{i\phi_-}$, or we have the necessary condition $\left|t\right|=\left|r\right|$ for CPA performance.

Since the high-order scatterings of the deep-subwavelength graphene nanoribbons are negligible, the graphene meta-surface can be formalized with effective surface conductivities ($\sigma_{\|}^e$) and its scatterings can be got with a transfer matrix formalism:
\begin{subequations}
\label{eq:whole}
\begin{equation}
t_\bot=\frac{2}{2+\sigma_{\|}^e \eta_0},
\end{equation}
\begin{equation}
r_\bot=\frac{\sigma_{\|}^e \eta_0}{2+\sigma_{\|}^e \eta_0},
\end{equation}
\end{subequations}
where $\eta_0$ is the wave impedance of free space.

It can be seen from Fig. 2(a) there exists two frequencies ($22.65$ THz, and $23.33$ THz), which we call \textit{quasi-CPA} point, where $\left| t\right| = \left| r\right|$ implies the necessary condition for suppressing the scattering fields to completely absorb coherent input beams of equal-intensity. In the view of experiments, graphene generally needs to be transferred onto some substrate, we studied the scattering responses of a nanoribbon array [with same geometry as that in Fig. 2(a)] sandwiched in between two $45$ nm-thick hexagonal boron nitride (h-BN) substrates, which was suggested as an exceptionally clean environment for achieving high confinement and low levels of plasmon damping in graphene\cite{Woessner2014} and is suitable for the one-dimensional high-quality electrical contact \cite{Wang2013} (see the illustration in Fig.1). The dielectric function of which was taken from experimental studies.\cite{Woessner2014,Caldwell2014} As can be seen from the inset of Fig. 2(a), the resonant frequency shifts to lower frequency as expected because of the introduction of the substrate, and there keeps two quasi-CPA frequencies.
\begin{figure}[t]
\includegraphics[width=7.2cm]{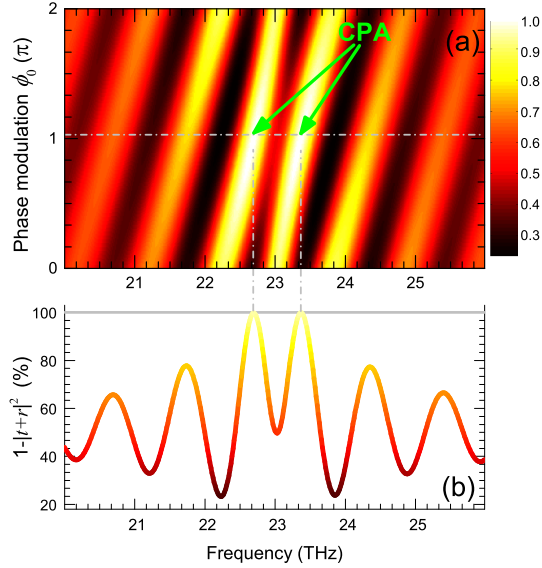}\caption{\label{fig3} A two-dimensional false-color plot of the normalized total absorptions as a function of frequency and phase modulation $\phi_0$, the exact CPA points are denoted with green arrows. (b) Normalized total absorption as a function of frequency for the phase modulation $\phi_0 = 1.03\pi$.}
\end{figure}
\begin{figure}[b]
\includegraphics[width=7.6cm]{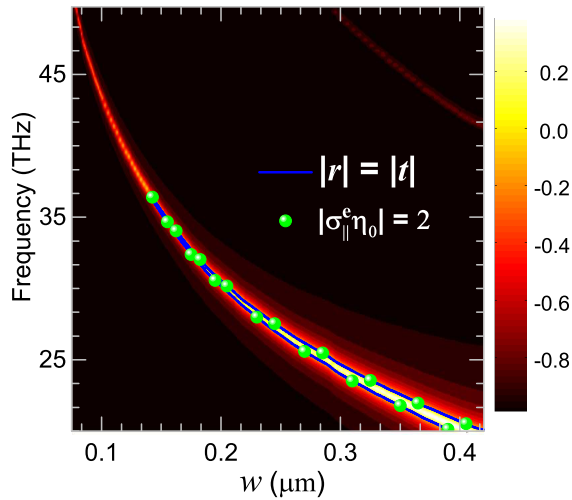}
\caption{\label{fig4} Geometric tunability of the graphene meta-surface CPA: Spectra of the difference ($\left|r\right|-\left|t\right|$) of the scattering coefficients ($r$ and $t$) for the graphene meta-surface width increasing from $0.075$ $\mu$m to $0.42$ $\mu$m. The solid line indicates quasi-CPA points where $\left|r\right| = \left|t\right|$, while the spheres represent the extracted surface conductivities with values $\left|\sigma_{\parallel}^{e}\eta_0\right| = 2$.}
\end{figure}

From Eq. 3, we can also get the formation condition for CPA in an effective medium scheme as: $\left|\sigma_{\|}^e\right| \eta_0 = 2$. To verify this, we used a recently proposed sheet retrieval method \cite{Fan2015,Tassin2012} to extract the effective surface conductivity $\sigma_{\parallel}^{e}$ of the graphene nanoribbon meta-surface. Figure 2(b) shows the absolute, real and imaginary parts of the effective surface conductivities corresponding to Fig. 2(a), It is obvious that there is a Lorentz-resonance around $23.0$ THz on the effective electric conductivity spectrum, which confirms the excitation of electric dipolar mode. The extracted magnetic conductivity does not show any resonant feature around this electric resonance, so we have not included the corresponding result here. And it is also seen that the condition $\left|\sigma_{\|}^e\right| \eta_0 = 2$ is fulfilled at the two quasi-CPA frequencies, which indicate the validity of describing the graphene nanoribbon meta-surface with the effective surface conductivity.

To implement the perfect absorption with the graphene meta-surface, we set a chirped phase modulation $\Delta \phi (f)=\phi_+ -\phi_- =\phi_0+kf$ on the beams $I_\pm$, with $f$ being the frequency, and $k=-1.87\times10^{-12}\pi$ being the chirped factor to compensate the frequency dispersion. The plotted false-color map of the normalized spectra of total absorptions in Fig. 3(a), shows the detailed dependence on $\phi_0$. We see that a proper phase modulation ($\phi_0=1.03\pi$) of the input coherent beams leads to significant suppression of the scattering outputs at the quasi-CPA frequencies. The normalized total absorption as a function of frequency for the phase modulation $\phi_0 = 1.03\pi$ is plotted in Fig. 3(b). We can see total absorption at both the two quasi-CPA frequencies. The significant boosting of the absorption implies destructive interference which prevents the coherent beams from escaping the absorbing channel of the graphene meta-surface, demonstrating completely absorption of the coherent input beams.

The meta-surface structures together with the electrically-controlled graphene will provide more wider tunable space for the design of mid-infrared CPA, we first consider the geometric tunability of the graphene nanoribbon based CPA. Figure 4 shows the dependence of the difference ($\left|r\right|-\left|t\right|$) of the scattering coefficients ($r$ and $t$) of the graphene meta-surface on the widths of nanoribbons. We can see that the resonant frequency of the electric dipolar mode shows a monotonous red-shift with the increase of $w$, which is similar the cut-wire case \cite{Fan2015} (actually, the ribbon structure is the special situation of cut-wire with graphene covers all the lattice range along $x$ axis. Increasing of the width $w$ or the graphene' filling factor in the unit cell of the meta-surface leads to stronger light-graphene interaction, i.e. high resonant strength of the electric dipolar resonance, and thus higher $r$ and lower $t$ around the resonance that introduce a regime where $\left|r\right|-\left|t\right| \geq 0$ starting from $w = 0.138$ $\mu$m, which has its boundary (as the solid line indicated) being the quasi-CPA points. The discrete spheres on top of the solid curve, representing the extracted surface conductivities with $\left|\sigma_{\parallel}^{e}\eta_0\right| = 2$, also imply the formation condition of CPA is fulfilled at the boundary. At these quasi-CPA points, we can completely suppress the scatterings of the graphene meta-furface with proper phase modulations as that showed in Fig. 3.
\begin{figure}[t]
\includegraphics[width=7.6cm]{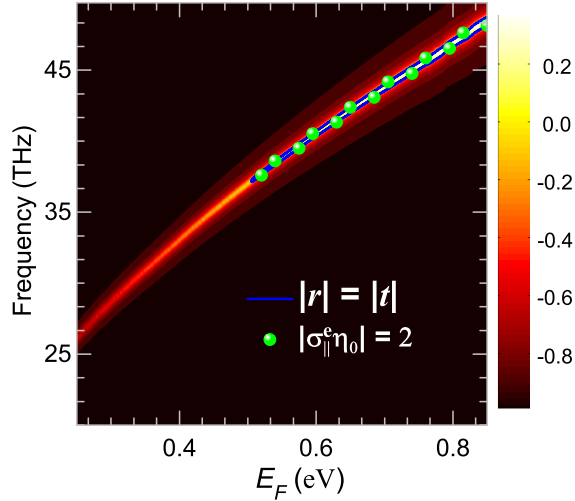}
\caption{\label{fig5}Electric tunability of the graphene meta-surface CPA: Spectra of the difference ($\left|r\right|-\left|t\right|$) of the scattering coefficients ($r$ and $t$) for the graphene meta-surface with Fermi energy $E_F$ ranging from $0.25$ eV to $0.85$ eV. The solid line indicates quasi-CPA points where $\left|r\right| = \left|t\right|$, while the spheres represent the extracted surface conductivities with values $\left|\sigma_{\parallel}^{e}\eta_0\right| = 2$.}
\end{figure}
\begin{figure}[t]
\includegraphics[width=8.6cm]{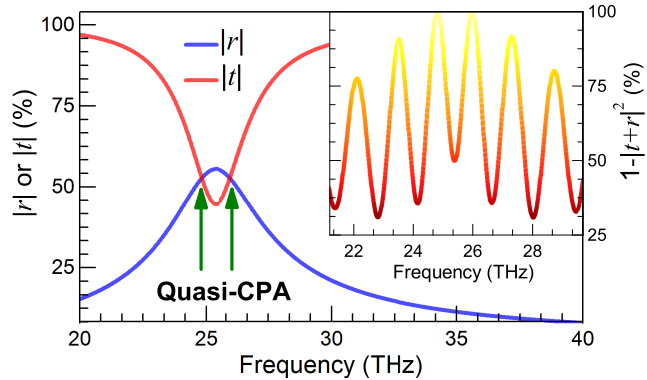}
\caption{\label{fig5}Scattering coefficients of a graphene nanoribbon array with experimental data (Yan \textit{et al.} data), the inset shows normalized total absorption for proper phase modulation.}
\end{figure}

On the other hand, the graphene meta-surface is also expected to have higher resonant strength for graphene with larger Fermi level.\cite{Abajo2014,Fan2015} The dependence on Fermi energy of the difference ($\left|r\right|-\left|t\right|$) of the scattering coefficients are plotted in Fig. 5 [the width of the graphene nanoribbon was set to be $0.138$ $\mu$m, the left edge of the solid boundary in Fig. 4], it can be seen that the resonant frequency shifts to higher frequencies and the resonant strength becomes higher while the Fermi level changes from to $0.25$ eV to $0.85$ eV. Similar to the influence of the width of nanoribbons, the light-graphene interaction is enhanced for better resonant behaviors, and thus a regime starting from $0.5$ eV where $\left|r\right|-\left|t\right| \geq 0$ with its boundary being the quasi-CPA frequencies. Combining the two functional bands for CPA, we see that it is freely to achieve CPA at desired frequency in an ultra-wide range by merging the geometric and electrical tunabilities. And the discrete spheres representing the extracted surface conductivities in Fig. 5 again confirm that $\left|\sigma_{\parallel}^{e}\eta_0\right| = 2$ is equivalent to the quasi-CPA condition $\left|r\right| = \left|t\right|$ for graphene nanoribbons based meta-surface.

Here, we also consider a graphene nanoribbon array ($P = 0.22$ $\mu$m, $w = 0.20$ $\mu$m) made from experimentally practical graphene, i.e., the Yan \textit{et al.} graphene with Drude weight $e^{2}E_{F}/\pi\hbar^{2} = 76.0$ GHz/$\Omega$ and collision frequency $\tau^{-1} = 9.8$ THz.\cite{Yan2012,Tassin2013} The spectra of the scattering coefficients are presented in Fig. 6, from which, a electric resonance is found around 25.5 THz with two associate quasi-CPA frequencies. The inset shows perfect absorption can be achieved at these two quasi-CPA frequencies with proper phase modulation. This suggest that the graphene nanoribbon meta-surface based CPA is feasible with high quality chemical vapour deposition (CVD) grown graphene samples.

In summary, we show that graphene nanoribbons based meta-surface can be employed for perfectly suppressing scattering of mid-infrared radiations for CPA. And the quasi-CPA frequency can be found from the effective surface conductivity at where $\left|\sigma_{\parallel}^{e}\eta_0\right| = 2$. Furthermore, the CPA can be tuned in a ultra-wide frequency band by considering both the geometric tunability and electrically-controlled charge-carrier density in graphene. We expect potential applications of coherent modulations in optical detections and signal processing with structured two-dimensional materials.

The authors would like to acknowledge financial support from the NSFC (Grants No. 11174221, 11204218, 11372248, 61101044, 61275176, 61390503, and 91323304), the Fundamental Research Funds for the Central Universities (Grant No. 3102015ZY079), and the National 863 Program of China (Grant No. 2012AA030403).

\bibliographystyle{aipnum4-1}
\nocite{*}

\bibliography{RibbonCPA}

\end{document}